\documentclass[aps,pra,preprintnumbers,amsmath,amssymb,floatfix,twocolumn]{revtex4-1}

\usepackage{graphicx}
\usepackage{indentfirst}
\usepackage{color,soul}
\usepackage{braket}
\usepackage{tabularx}
\usepackage{hyperref}

\begin{document}
\title{Quantum illumination receiver using double homodyne detection}

\author{Yonggi Jo}\email[]{yonggi@add.re.kr}
\affiliation{Quantum Physics Technology Directorate, Advanced Defense Technology Research Institute, Agency for Defense Development, Yuseong P.O. Box 35,  Daejeon 34186, Republic of Korea}
% ORCID : https://orcid.org/0000-0002-2992-304X

\author{Sangkyung Lee}
\affiliation{Quantum Physics Technology Directorate, Advanced Defense Technology Research Institute, Agency for Defense Development, Yuseong P.O. Box 35, Daejeon 34186, Republic of Korea}

\author{Yong Sup Ihn}
\affiliation{Quantum Physics Technology Directorate, Advanced Defense Technology Research Institute, Agency for Defense Development, Yuseong P.O. Box 35, Daejeon 34186, Republic of Korea}

\author{Zaeill Kim}
\affiliation{Quantum Physics Technology Directorate, Advanced Defense Technology Research Institute, Agency for Defense Development, Yuseong P.O. Box 35, Daejeon 34186, Republic of Korea}

\author{Su-Yong Lee}\email[]{suyong2@add.re.kr}
\affiliation{Quantum Physics Technology Directorate, Advanced Defense Technology Research Institute, Agency for Defense Development, Yuseong P.O. Box 35, Daejeon 34186, Republic of Korea}

\date{\today}

\begin{abstract}
A quantum receiver is an essential element of quantum illumination (QI) which outperforms its classical counterpart, called classical-illumination (CI). However, there are only few proposals for realizable quantum receiver, which exploits nonlinear effects leading to increasing the complexity of receiver setups. To compensate this, in this article, we design a quantum receiver with linear optical elements for Gaussian QI. Rather than exploiting nonlinear effect, our receiver consists of a 50:50 beam splitter and homodyne detection. Using double homodyne detection after the 50:50 beam splitter, we analyze the performance of the QI in different regimes of target reflectivity, source power, and noise level. We show that our receiver has better signal-to-noise ratio and more robust against noise than the existing simple-structured receivers.
\end{abstract}

\maketitle

\section{Introduction}

Superposition and entanglement are properties mainly exploited in quantum information processing protocols, such as quantum communication \cite{Bennett1984, Ekert1991} and quantum computing \cite{Feynman1982}. In the protocols, it is very crucial issue protecting these quantum mechanical phenomena during the process, since they are very fragile against decoherence. In 2008, S. Lloyd presented a binary hypothesis testing protocol using entangled states in a single-photon level, called quantum illumination (QI), to improve a capability of target detection in an optical radar \cite{Lloyd2008}. Different from other quantum information processing protocols, it was shown that QI has advantages compared with its classical counterpart, called classical-illumination (CI), with the same transmission energy under a decoherence channel, even when entanglement is not left after passing through the channel.

After its first proposal, there have been many studies about QI \cite{Tan2008, Guha2009,SL09, Wilde2017, Lopaeva2013, Zhang2015, England2019, Luong2019,Barzanjeh2015, Barzanjeh2020, Karsa2020, Devi, Ragy, Zhang14, Sanz, Liu17, Weedbrook, Bradshaw, Zubairy, Stefano19, Palma, Ray, Sun, Ranjith, Sandbo, Aguilar, Sussman, Lee, Zhuang2017, Karsa2020-1,Yung,Shapiro2019,Guha2009-1}. Since thermal noise baths and an optical entangled state generated from spontaneous parametric down-conversion (SPDC) can be written in Gaussian state form, it is more realistic to study Gaussian QI\cite{Tan2008,SL09,Wilde2017,Karsa2020}. Under a very noisy channel, it was shown that Gaussian QI system outperforms the optimal CI exploiting a coherent state transmitter under the same transmission energy. The QI was experimentally demonstrated in laboratories \cite{Lopaeva2013,Zhang2015,England2019}. Furthermore, to exploit more appropriate spectral region for a target detection protocol than optical wavelengths, microwave QI was studied \cite{Barzanjeh2015} and demonstrated \cite{Luong2019,Barzanjeh2020} as well.

In the previous QI studies, it was shown whether the presence or absence of a target with very low reflectivity can be more precisely discriminated using an entangled state than a coherent state. The precision limit is determined by an error probability of the hypothesis test problem, and it is upper bounded by the quantum Chernoff bound \cite{Audenaert2007,Calsamiglia2008,Pirandola2008}. Given a probe state in a channel, we can derive the quantum Chernoff bound which is accompanied with the corresponding optimal measurement setup.

There are few studies about quantum receivers for QI which are sub-optimal while outperforming the CI. Guha and Erkmen presented the optical parametric amplifier (OPA) receiver and the phase conjugate (PC) receiver \cite{Guha2009,Guha2009-1} which were experimentally demonstrated at optical frequency \cite{Zhang2015} and at microwave domain \cite{Barzanjeh2020}, respectively. A scheme of feed-forward sum-frequency generation (FF-SFG) \cite{Zhuang2017} asymptotically approaches to the quantum Chernoff bound, but it has not been demonstrated due to the hardness of its implementation. Those quantum receivers are designed for exploiting nonlinear effects in order to measure correlation between two modes used in QI. By using the nonlinear effects, a QI system with one of these receivers can outperform a CI system in the hypothesis testing problem. However, many incoming signals which do not interact with nonlinear media are discarded, such that the inefficiency of the nonlinear effect diminishes signal-to-noise ratio (SNR) of the entire QI system.

In this article, we propose a quantum receiver for Gaussian QI which does not include a nonlinear optical element. Our setup is constructed with a 50:50 beam splitter and homodyne detection which is widely used in quantum information processing with continuous variables. Because of the absence of nonlinear effect, our setup is simple to implement compared with other receivers. Since a Gaussian state exploited in QI has zero-mean, mean-square values derived by homodyne detection, i.e., the second-order moments of the Gaussian state, are used to discriminate the two hypotheses. We investigate the error probabilities of a QI system with various receivers, choosing the best receiver among the three receivers in various target reflectivity, source power, and noise level, while the existing studies about a QI receiver considered a fixed condition \cite{Guha2009,Guha2009-1,Zhuang2017}. We show our receiver is suitable for Gaussian QI with low energy source in a very noisy channel than the OPA and PC receiver. Also, we analyze SNR of a QI system with our receiver, which can show better SNR than the OPA and PC receivers.

This article is organized as follow. In Sec.~\ref{SecQI} and Sec.~\ref{SecSNR}, we introduce basic concepts of QI and tools for analyzing performance of QI, respectively. In Sec.~\ref{SecSHD}, we propose a receiver setup which is constructed with a 50:50 beam splitter and homodyne detection. In Sec.~\ref{SecAnalysis}, we analyze the performance of QI with various receivers in different conditions. Finally, it is summarized with discussion in Sec.~\ref{SecCon}.

\section{Quantum illumination}\label{SecQI}

\begin{figure}[t!]
	\includegraphics[width=0.36\textwidth]{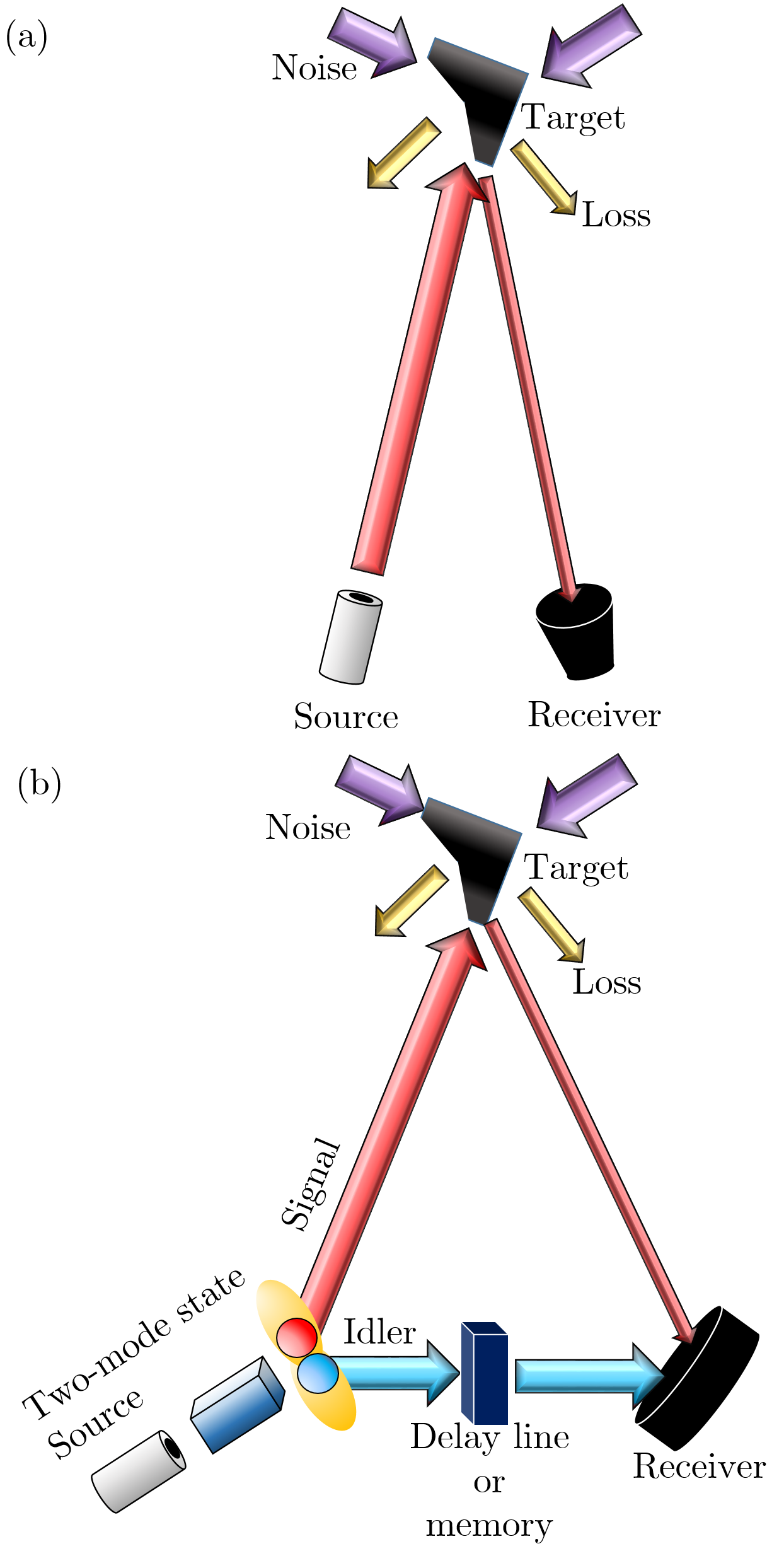}
	\caption{Schematic diagrams of CI and QI. (a) In CI, a signal beam, is sent to the target, and the return beam is measured to discriminate the presence and absence of the target. (b) In QI, an entangled state is exploited to discriminate the two situations. The signal beam is sent to the target, and the return beam is jointly measured with the idler beam. By using joint measurement of the two beams, QI can have better efficiency for the hypothesis test than CI.}\label{FigCIQI}
\end{figure}

The purpose of a target detection protocol is to discriminate following two situations: a target is absent (hypothesis $H_{0}$) or present (hypothesis $H_{1}$). In CI, a signal is sent to the target, and the return signal is measured to discriminate the two situations as shown in Fig.~\ref{FigCIQI} (a). In QI, an entangled state is exploited, as shown in Fig.~\ref{FigCIQI} (b). QI takes advantages over CI to detect a target in a lossy and noisy channel \cite{Lloyd2008}. After this study, QI described with Gaussian state, called Gaussian QI was studied \cite{Tan2008}. The Gaussian QI can provide more realistic and exact statistics than the original one since thermal noise baths are in Gaussian regime under Bose-Einstein statistics and the entangled beams generated from continuous wave SPDC are described with Gaussian states, e.g., a two-mode squeezed vacuum (TMSV) state.

A TMSV state can be expressed in the photon number basis as follows:
\begin{align}
	\ket{\text{TMSV}}=\sum_{n=0}^{\infty}\sqrt{\frac{N_{S}^{n}}{(N_{S}+1)^{n+1}}}\ket{n}_{S}\ket{n}_{I},
\end{align}
where $N_{S}$ is the mean photon number per each mode, and the subscripts $S$ and $I$ denote signal and idler modes. For calculation of Gaussian states, it is convenient to describe the state in quadrature representation. Since a TMSV state has zero-mean, its covariance matrix can be written as follows:
\begin{align}
	V_{\text{TMSV}} = \begin{pmatrix}
		A & 0 & C & 0 \\
		0 & A & 0 & -C \\
		C & 0 & A & 0\\
		0 & -C & 0 & A
	\end{pmatrix},
\end{align}
where $A=2N_{S}+1$, $C=2\sqrt{N_{S}(N_{S}+1)}$.

\begin{figure}[t!]
	\includegraphics[width=0.4\textwidth]{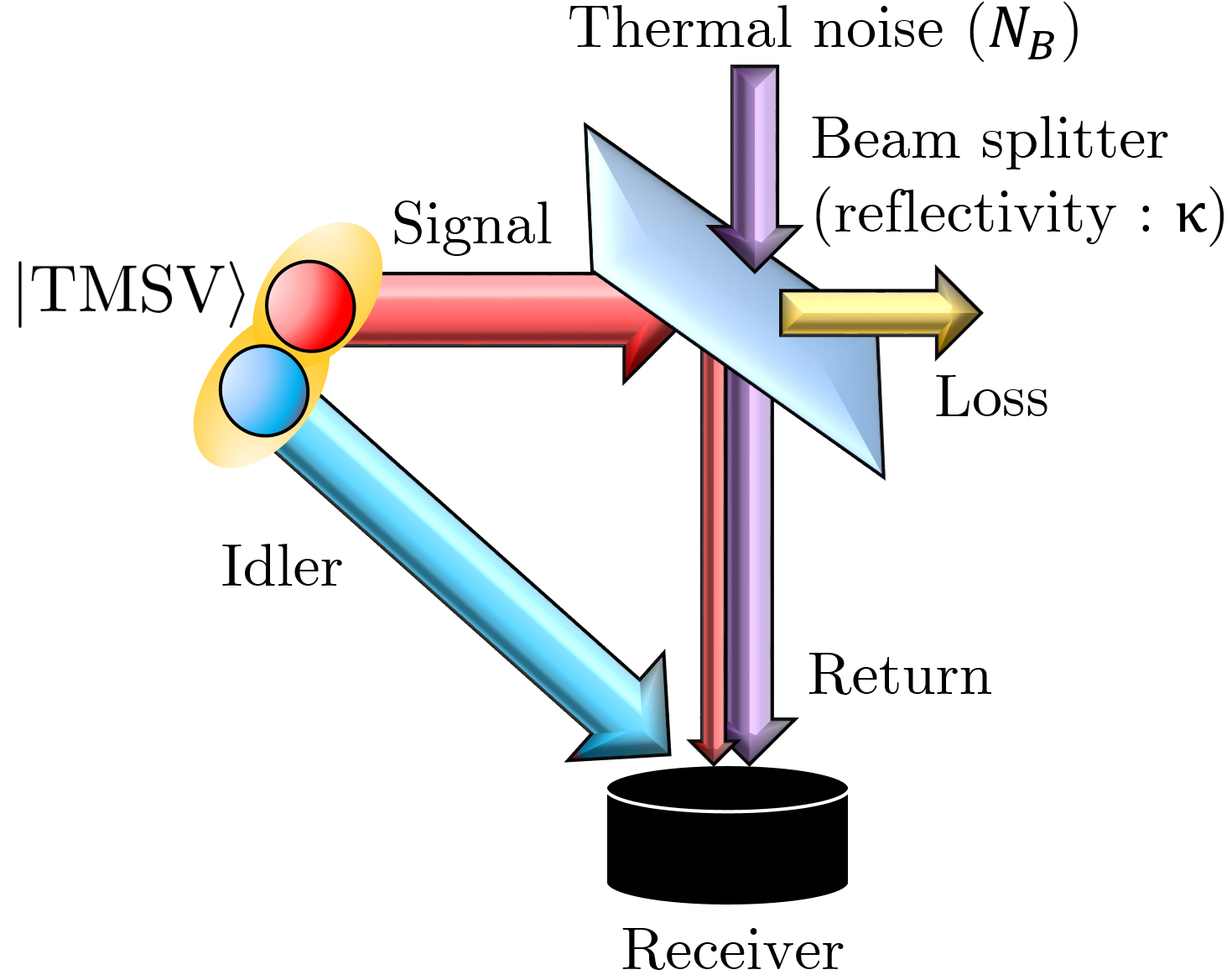}
	\caption{A schematic diagram of proof-of-principle model for QI. The signal beam of a TMSV state propagates to the beam splitter whose reflectivity is $\kappa$. At the beam splitter, thermal noise is induced, and the return beam and the idler beam are jointly measured at the receiver.}\label{FigQIM}
\end{figure}

In Gaussian QI, a signal is sent to a target with very low reflectivity in a thermal noise background, and the idler is kept intact. A schematic diagram of the proof-of-principle QI model is drawn in Fig.~\ref{FigQIM}. A target with low reflectivity is realized with a beam splitter with low reflectivity, and the signal is combined with a thermal noise at the beam splitter. Finally, the return and idler beams are jointly measured at a receiver. Under the hypothesis $H_{0}$, the return mode annihilation operator will be $\hat{a}_{R}=\hat{a}_{B}$, where $\hat{a}_{B}$ is the annihilation operator of thermal noise with the mean photon number $N_{B}$. Under the hypothesis $H_{1}$, the return mode annihilation operator will be $\hat{a}_{R}=\sqrt{\kappa}\hat{a}_{S}+\sqrt{1-\kappa}\hat{a}_{B}$, where $\hat{a}_{S}$ is the annihilation operator of the signal mode. The covariance matrix of the return mode and the idler mode under the hypothesis $H_{0}$ is:
\begin{align}\label{H0}
	V_{0}=\begin{pmatrix}
		B & 0 & 0 & 0\\
		0 & B & 0 & 0\\
		0 & 0 & A & 0\\
		0 & 0 & 0 & A
	\end{pmatrix},
\end{align}
where $B=2N_{B}+1$. Since there is no correlation between the return mode and the idler mode, Eq.~(\ref{H0}) has null off-diagonal terms. The covariance matrix under the hypothesis $H_{1}$ is:
\begin{align}\label{H1}
	V_{1}=\begin{pmatrix}
		\kappa A+(1-\kappa)B & 0 & C\sqrt{\kappa} & 0\\
		0 & \kappa A+(1-\kappa)B & 0 & -C\sqrt{\kappa}\\
		C\sqrt{\kappa} & 0 & A & 0\\
		0 & -C\sqrt{\kappa} & 0 & A
	\end{pmatrix},
\end{align}
which contains non-null off-diagonal terms since there is correlation between the return mode and the idler mode.

From the two covariance matrices, one discriminates the two hypotheses based on the off-diagonal terms. To obtain the off-diagonal elements from measurement results, it is necessary to interfere the return mode with the idler mode before the measurement. There are few studies about a QI receiver, such as OPA receiver, PC receiver \cite{Guha2009,Guha2009-1,Karsa2020-1}, and FF-SFG receiver \cite{Zhuang2017}. To interact the two modes, those receivers contain nonlinear optical elements leading to increasing the complexity of receiver setups. Therefore, it seems to be necessary to construct a QI receiver which is simply implemented with high SNR, excluding nonlinear effects.

\section{Signal-to-noise ratio\\in quantum illumination}\label{SecSNR}

In this section, we introduce the calculation of SNR for Gaussian QI, which evaluates the performance of a Gaussian QI system. Before we explain the calculation of SNR, we defined the following notations for simplification:
\begin{align}\label{EXPV}
	\begin{split}
		R_{x}=\text{Tr}\left(\hat{M}\hat{\rho}_{x}\right),\quad
		\Delta R_{x}=\text{Tr}\left(\hat{M}^{2}\hat{\rho}_{x}\right)-\left[\text{Tr}\left(\hat{M}\hat{\rho}_{x}\right)\right]^{2},
	\end{split}
\end{align}
where $\hat{\rho}_{x}$ denotes the density matrix corresponding to the hypothesis $H_{x}$ and $\hat{M}$ denotes a measurement operator. For a given density matrix, the two equations in Eq.~(\ref{EXPV}) denote the expectation value and the variance of a given measurement operator. If we exploit $K$-mode rather than a single mode, the expectation value and the variance become $K R_{x}$ and $K\Delta R_{x}$, respectively.

SNR is derived from error probabilities of decision problem. For a binary hypothesis test, a threshold $R_{\text{Th}}$ should be defined for a problem. Then, the problem can be decided based on this threshold. For example, the hypothesis $H_{0}$ is considered as true when a result of Gaussian QI is above $R_{\text{Th}}$, and the hypothesis $H_{1}$ is true when the result is below $R_{\text{Th}}$. However, there can be an error in the decision, such as a false alarm or a miss detection. The false alarm is the case that the decision is target presence even if there is no target, and the miss detection means the case that the decision is target absence even when the target presents. When the two hypotheses are equally probable, the total error probability can be written as follows:
\begin{align}
	P_{E}=\frac{1}{2}P(1|0)+\frac{1}{2}P(0|1),
\end{align}
where $P(1|0)$ means the false alarm probability, and $P(0|1)$ does the miss detection probability. According to the above description, the decision will be target presence for $<R_{\text{Th}}$ and target absence for otherwise. 
Here, we consider a large number of independent signal-idler mode pairs, i.e, $K\gg 1$. Due to the central limit theorem, the error probabilities approach Gaussian distributions of which mean and variance are $K R_{x}$ and $K\Delta R_{x}$, respectively \cite{Guha2009}. Each error probability can be calculated from the following equations:
\begin{align}
	\begin{split}
		P(1|0)&=\int_{-\infty}^{R_{\text{Th}}}\frac{dx}{\sqrt{2\pi K \Delta R_{0}}}\exp\left[-\frac{1}{2}\left(\frac{x-K R_{0}}{\sqrt{K \Delta R_{0}}}\right)^2\right],\\
		P(0|1)&=\int_{R_{\text{Th}}}^{\infty}\frac{dx}{\sqrt{2\pi K \Delta R_{1}}}\exp\left[-\frac{1}{2}\left(\frac{x-K R_{1}}{\sqrt{K \Delta R_{1}}}\right)^2\right],
	\end{split}
\end{align}
and the results are
\begin{align}
	\begin{split}
		P(1|0)&=\frac{1}{2}\text{erfc}\left[\frac{K R_{0}-R_{\text{Th}}}{\sqrt{2 K \Delta R_{0}}}\right],\\
		P(0|1)&=\frac{1}{2}\text{erfc}\left[\frac{R_{\text{Th}}-K R_{1}}{\sqrt{2 K \Delta R_{1}}}\right].
	\end{split}
\end{align}
From the definition of the complementary error function and the relation $R_{1}\leq R_{\text{Th}}\leq R_{0}$, $P(1|0)$ and $P(0|1)$ are in the trade-off relation about $R_{\text{Th}}$, and the total error probability is minimized when the two error probabilities are the same. Thus, the threshold of decision which minimizes the total error probability is obtained from the following equation:
\begin{align}\label{TM}
	R_{\text{Th}}=\frac{K(R_{0}\sqrt{\Delta R_{1}}+R_{1}\sqrt{\Delta R_{0}})}{\sqrt{\Delta R_{0}}+\sqrt{\Delta R_{1}}}.
\end{align}
With the threshold, the total error probability can be calculated from the following equation \cite{Guha2009,Guha2009-1}:
\begin{align}\label{SNRD}
	\begin{split}
		P_{\text{E}}&=\frac{1}{2}\text{erfc}\left[\frac{\sqrt{K}(R_{0}-R_{1})}{\sqrt{2}(\sqrt{\Delta R_{0}}+\sqrt{\Delta R_{1}})}\right]\\
		&\approx \frac{\exp[-\text{SNR}^{(K)}]}{2\sqrt{\pi \text{SNR}^{(K)}}},
	\end{split}
\end{align}
where $\text{SNR}^{(K)}$ is an SNR and it is defined as follows:
\begin{align}\label{SNRdef}
	\text{SNR}^{(K)}\equiv\frac{K\left(R_{0}-R_{1}\right)^{2}}{2\left(\sqrt{\Delta R_{0}}+\sqrt{\Delta R_{1}}\right)^{2}},
\end{align}
which is a conventional squared SNR equation when background bias exists \cite{Zhang2015,Barzanjeh2020}. The approximation in Eq.~(\ref{SNRD}) is true only when $\text{SNR}^{(K)}\gg 1$. From Eq.~(\ref{SNRD}), we find that the error probability becomes lower with increasing $\text{SNR}^{(K)}$, i.e., the higher SNR means the more accurate decision in Gaussian QI.

\section{Double homodyne detection}\label{SecSHD}

\begin{figure}[t!]
	\includegraphics[width=0.47\textwidth]{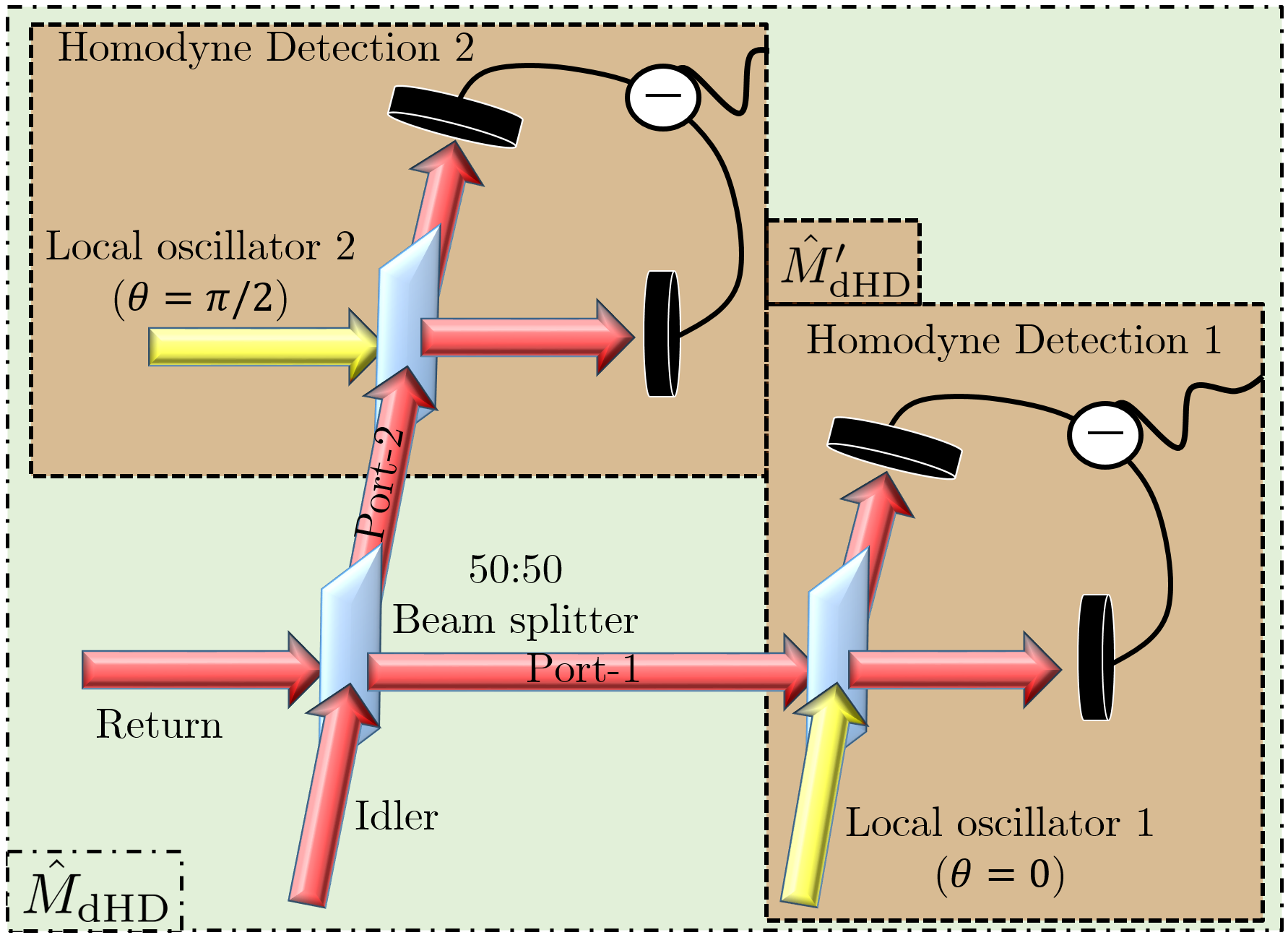}
	\caption{A schematic diagram of the double HD. A return beam and an idler beam are combined at 50:50 beam splitter, and then, double HD are performed on each output port of the beam splitter. In the port-1, HD on the position ($\theta=0$) is performed, and HD on the momentum ($\theta=\pi/2$) is performed in the other output port. The setup in dashed lines denotes $\hat{M}_{\text{dHD}}'$ expressed in Eq.~(\ref{mprime}), and the measurement operator of the whole setup in dot-dashed lines is $\hat{M}_{\text{dHD}}$ written in Eq.~(\ref{SHD}).}\label{FigSHD}
\end{figure}

\begin{figure*}[t!]
	\includegraphics[width=0.7\textwidth]{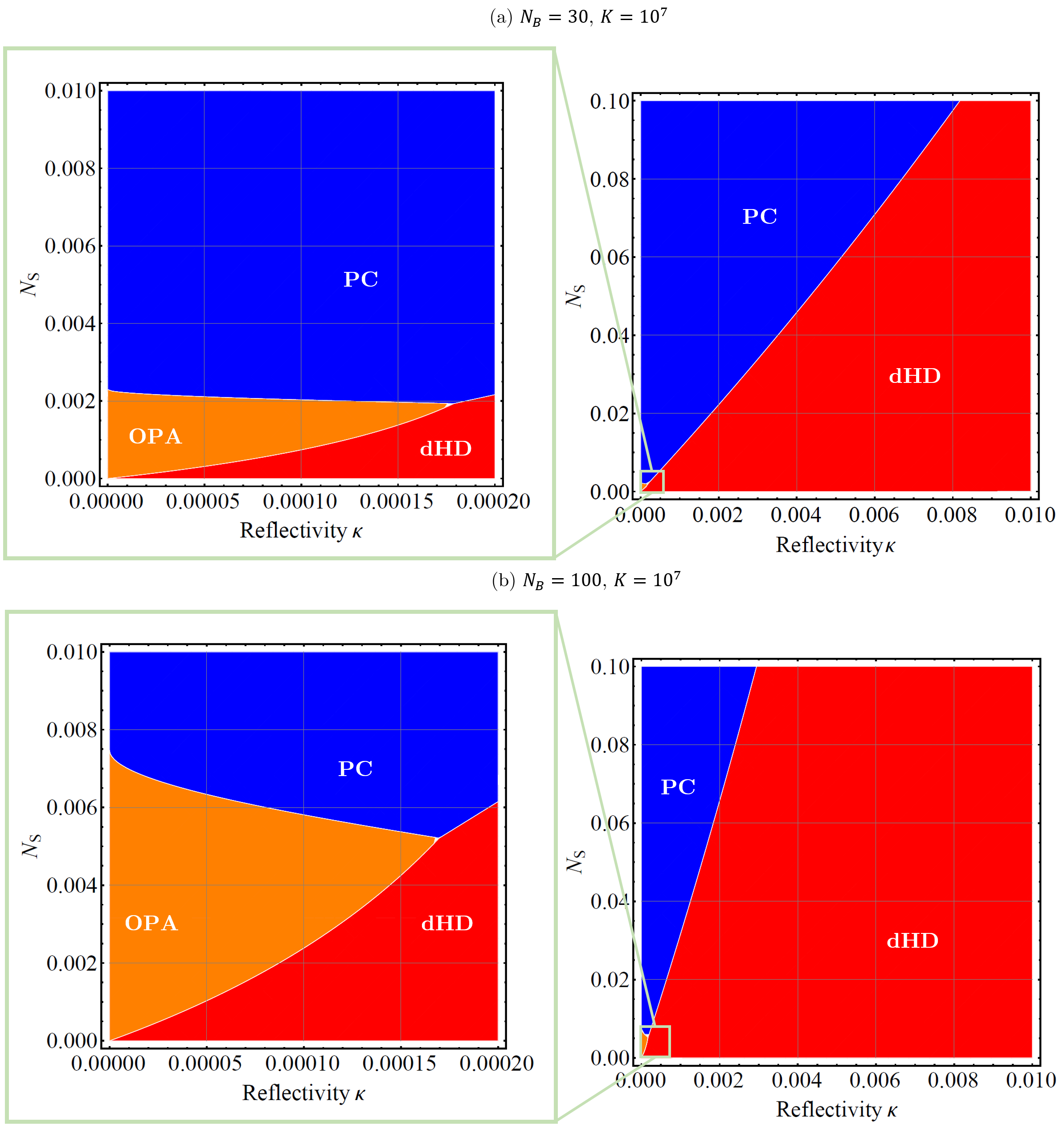}
	\caption{The regions, where one of the receivers outperforms the others, are shown with various target reflectivity $\kappa$ and mean photon number of the signal $N_{S}$. Solid lines represent the boundaries that the two receivers provide the same SNR when they are adapted in a Gaussian QI system. (a) The mean photon number of thermal noise $N_{B}$ is $30$, and the number of modes $K$ is $10^{7}$. (b) The case that $N_{B}=100$ and $K=10^{7}$. The region that the double HD outpereforms the other receivers becomes wider. ``dHD" is abbreviation of ``double HD".}\label{FigSNRNS}
\end{figure*}

In this section, we describe the double HD as a tool of our QI receiver. We denote a balanced HD as a HD which consists of a 50:50 beam splitter, local oscillator (strong LASER) of which intensity is at least 10,000 times larger than the input signal one \cite{Braunstein1990}, and two intensity detectors. An input signal and a local osillator are coherently impinged on the 50:50 beam splitter. Then we measure the intensity difference between the output ports repeatedly, resulting in an expectation value of a quadrature operator as follows:
\begin{align}
	\braket{\hat{n}_{a}-\hat{n}_{b}}=|\alpha_{L}|\braket{\hat{X}(\theta)}=|\alpha_{L}|\left<\frac{\hat{a}^{\dagger}e^{i\theta}+\hat{a}e^{-i\theta}}{\sqrt{2}}\right>,
\end{align}
where $\hat{n}$ represents a number operator, $\hat{a}$ and $\hat{a}^{\dagger}$ denote annihilation and creation operators of the input mode, and the subscripts $a$, $b$ are labels of each intensity detector. $\theta$ is controlled by a phase of the local oscillator, and $\alpha_{L}$ is the amplitude of the local oscillator. If $\theta=0$, the HD setup measures the position of the input mode, and if $\theta=\pi/2$, it becomes momentum measurement.

The schematic diagram of the double HD setup is described in Fig.~\ref{FigSHD}. The idler and return beams are mixed by using a 50:50 beam splitter. Subsequently we perform HD on each output port, which we call double HD. One of the HD setups measures postion ($\theta=0$), and the other does momentum ($\theta=\pi/2$). Since zero-mean Gaussian states are exploited in our Gaussian QI, the expectation value of the quadrature operator is always zero. 
To obtain a non-zero expectation value, we use squared outcomes of homodyne detection. Then, we obtain an expectation value of a square quadrature operator $\braket{\hat{X}^{2}(\theta)}=\int_{-\infty}^{\infty}dx x^{2}P(x,\theta)$, where the phase rotated probability distribution $P(x,\theta)$ is obtained with $\braket{\hat{X}(\theta)}$ by repeated measurements. The square quadrature operators by the double HD can be written as:
\begin{align}\label{mprime}
	\hat{M}_{\text{dHD}}'=\left[\hat{X}_{1}(0)\right]^{2}+\left[\hat{X}_{2}\left(\pi/2\right)\right]^{2},
\end{align}
where the subscripts $1$ and $2$ denote labels of the output port of the 50:50 beam splitter, as shown in Fig.~\ref{FigSHD}. By taking a reverse 50:50 beam splitting operation, we can transform Eq.~(\ref{mprime}) into
\begin{align}\label{SHD}
	\begin{split}
		\hat{M}_{\text{dHD}}=&\hat{U}_{BS}\hat{M}_{\text{dHD}}'\hat{U}_{BS}^{\dagger}\\
		=&\hat{a}_{R}\hat{a}_{R}^{\dagger}-\hat{a}^{\dagger}_{R}\hat{a}^{\dagger}_{I}-\hat{a}_{R}\hat{a}_{I}+\hat{a}_{I}^{\dagger}\hat{a}_{I},
	\end{split}
\end{align}
where $\hat{U}_{BS}$ is the 50:50 beam splitting operator which is described in the following equation:
\begin{align}
	\begin{pmatrix}
		\hat{a}_{1} \\
		\hat{a}_{2}
	\end{pmatrix}=
		\hat{U}_{BS}\begin{pmatrix}
			\hat{a}_{R} \\
			\hat{a}_{I}
		\end{pmatrix}=\frac{1}{\sqrt{2}}\begin{pmatrix}
		1 & 1\\
		-1 & 1
		\end{pmatrix}
	\begin{pmatrix}
		\hat{a}_{R} \\
		\hat{a}_{I}
	\end{pmatrix}.
\end{align}
The measurement operator in Eq.~(\ref{SHD}) includes phase-sensitive cross-correlation components, $\hat{a}_{R}^{\dagger}\hat{a}_{I}^{\dagger}+\hat{a}_{R}\hat{a}_{I}$, of which expectation value gives the off-diagonal term in the covariance matrices written in Eqs.~(\ref{H0}) and (\ref{H1}). Consequently, we constructed the measurement setup which can obtain a correlation between the return and idler modes by using a beam splitter and double HD, rather than by exploiting nonlinear optical elements.

The double HD operator of Eq.~(\ref{SHD}) is compared to the measurement operators of the OPA and PC receivers. The measurement operator of the OPA receiver is written in the following equation:
\begin{align}\label{OPAM}
\begin{split}
\hat{M}_{\text{OPA}}=&(G-1)\hat{a}_{R}\hat{a}_{R}^{\dagger}\\
&+\sqrt{G(G-1)}(\hat{a}^{\dagger}_{R}\hat{a}^{\dagger}_{I}+\hat{a}_{R}\hat{a}_{I})+G\hat{a}_{I}^{\dagger}\hat{a}_{I},
\end{split}
\end{align}
where $G$ is a gain of the OPA $(G>1)$. The measurement operator of the PC receiver is:
\begin{align}\label{PCM}
\hat{M}_{\text{PC}}=\nu (\hat{a}_{R}\hat{a}_{I}+\hat{a}_{R}^{\dagger}\hat{a}_{I}^{\dagger})+\mu(\hat{a}_{I}\hat{a}_{V}^{\dagger}+\hat{a}_{I}^{\dagger}\hat{a}_{V}).
\end{align}
where $\hat{a}_{V}$ is a vacuum state operator, and $|\mu|^{2}-|\nu|^{2}=1$. Both receivers are constructed in order to measure phase-sensitive cross-correlation components, $\hat{a}_{R}^{\dagger}\hat{a}_{I}^{\dagger}+\hat{a}_{R}\hat{a}_{I}$. 
To compare our receiver with feasible receivers, we choose the parameter values of the receivers which are experimentally given or implementable. For the OPA receiver, the gain $G$ is experimentally obtained as $G-1=7.4\times 10^{-5}$\cite{Zhang2015}. 
For the PC receiver, the parameter values are implementable as $\mu=\sqrt{2}$ and $\nu=1$\cite{Guha2009, Guha2009-1}.
%In our analysis, we assign $G-1=7.4\times 10^{-5}$ for the OPA receiver and $\mu=\sqrt{2}$ and $\nu=1$ for the PC receiver. This gain $G$ of the OPA was used in the demonstration of QI \cite{Zhang2015}, and the coefficients of the PC receiver follow the values in the paper that the PC receiver is originally proposed \cite{Guha2009, Guha2009-1}, and these values were used in the demonstration as well \cite{Barzanjeh2020}. 

In the viewpoint of the measurement operators, we simply infer that our double HD operator can provide us a higher SNR than the operators of the OPA and PC receivers. There are two reasons as follows: First, the coefficients of phase-sensitive cross-correlation components are comparable to ones of the other components in Eq.~(\ref{SHD}), whereas they are smaller than the other terms in the Eqs.~(\ref{OPAM}) and (\ref{PCM}). Second, the coefficient of the return mode component $\hat{a}_{R}\hat{a}_{R}^{\dagger}$ is also comparable to the others in Eq.~(\ref{SHD}), whereas it is very small in Eq.~(\ref{OPAM}) and does not exist in Eq.~(\ref{PCM}). Based on the intuitive view, we observe how the measurement operators work out in the next section.

\section{Signal-to-noise ratio analysis}\label{SecAnalysis}

\begin{figure}[t!]
	\includegraphics[width=0.47\textwidth]{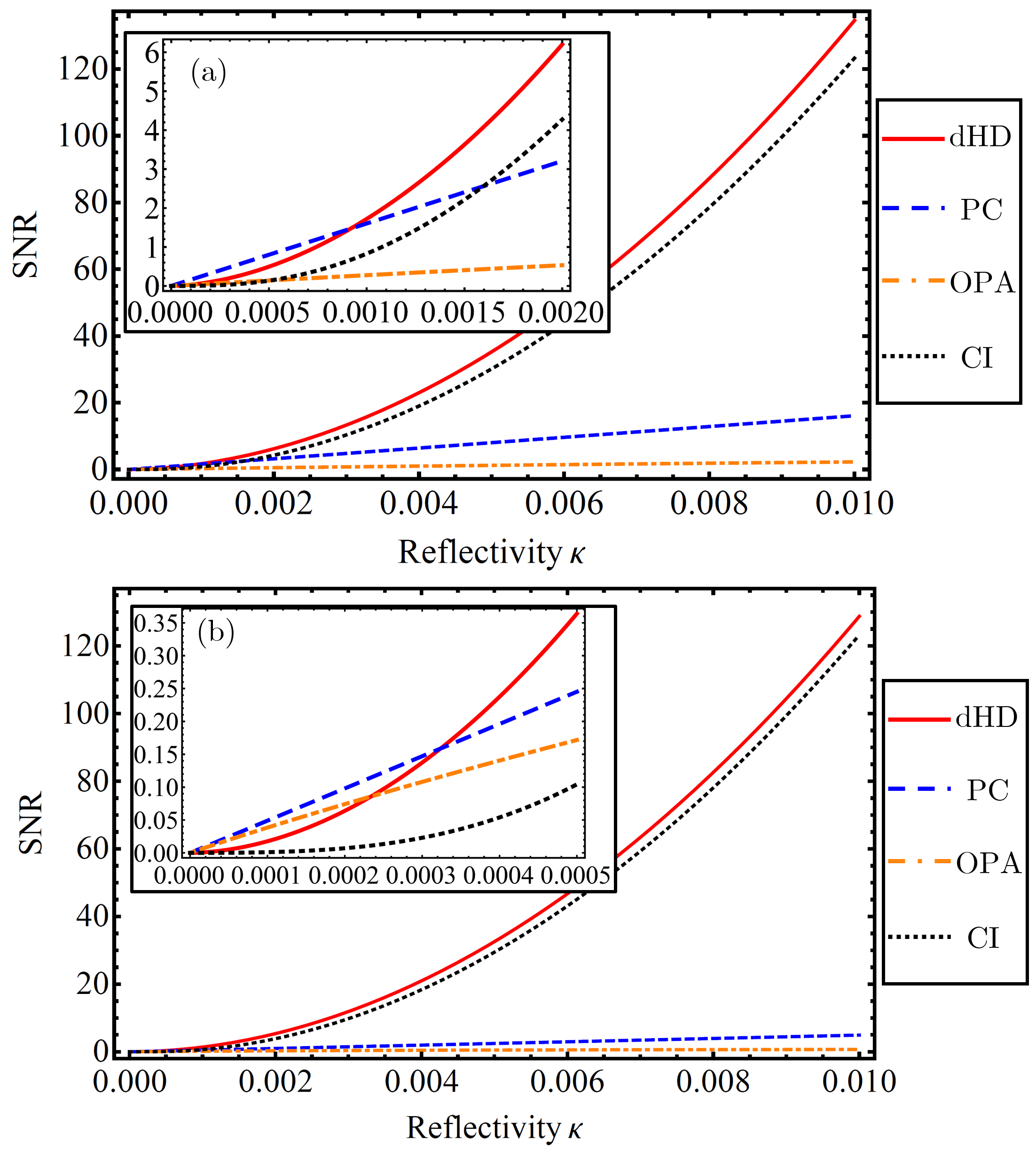}
	\caption{The plots of SNR$^{(K)}$ of the double HD (red solid lines), the PC receiver (blue dashed lines), the OPA receiver (orange dot-dased lines), and the coherent state CI (black dotted lines) at $N_{S}=0.01$ and $K=10^{7}$. The plots (a) and (b) show the SNR with $N_{B}=30$ and $100$, respectively.}\label{FigSNRD}
\end{figure}

\begin{figure*}[t!]
	\includegraphics[width=0.95\textwidth]{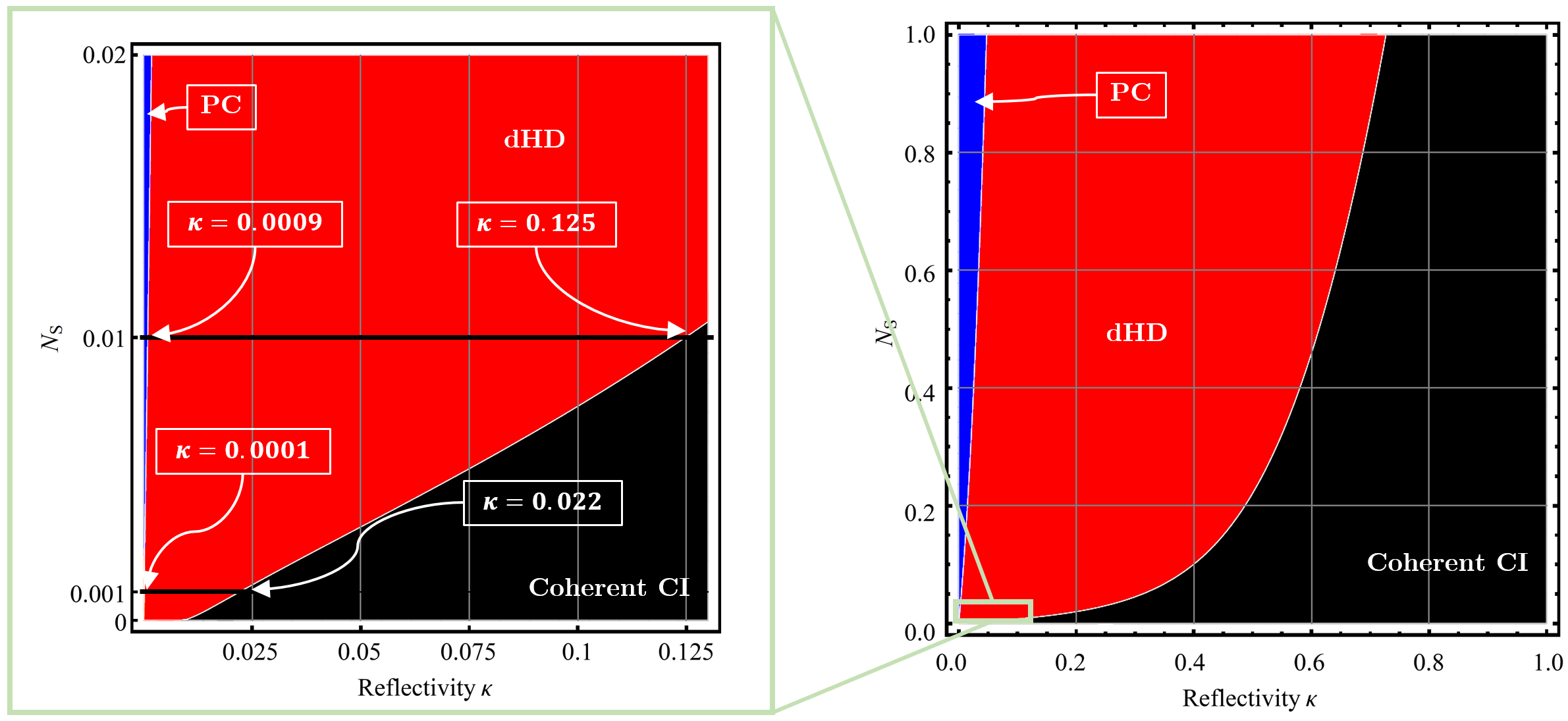}
	\caption{The regions, where one of the receivers outperforms the others, are shown with various target reflectivity $\kappa$ and mean photon number of the signal $N_{S}$ at $N_{B}=30$ and $K=10^{7}$. The two black horizontal lines denote the cases, $N_{S}=0.01$ and $N_{S}=0.001$.}\label{FigSNRall}
\end{figure*}

Using our double HD setup, we investigate the performance of Gaussian QI in different regimes of target reflectivity and source power, which is compared with the OPA receiver and the PC receiver \cite{Guha2009,Guha2009-1}. 
The detail expressions of the corresponding SNRs are given in appendix.

In Fig.~\ref{FigSNRNS}, the regions, where one of the receivers outperforms the others, are shown with various target reflectivity and mean photon number of the signal $N_{S}$. The regions are plotted based on the SNR of a QI system using each receiver. As a benchmark, SNR of a CI system is considered with a coherent state having a mean photon number $N_{S}$. The SNR is related with the quantum Chernoff bound which represents an upper bound of the error probability of a quantum discrimination problem for a given signal and channel \cite{Audenaert2007, Calsamiglia2008, Pirandola2008}. Thus, to claim that the QI system takes advantages over the CI, the SNR of a QI system should be higher than that of the CI.

Fig.~\ref{FigSNRNS} (a) shows the region plot when the mean photon number of thermal noise $N_{B}$ is $30$ and the number of exploited modes is $10^{7}$. Our receiver, the double HD, can outperform the other receivers at low reflectivity when $N_{S}$ is very small, i.e., our receiver is the most suitable for a QI system with a low-power source. The OPA and PC receivers need stronger signal for target detection due to an efficiency of the nonlinear optical effects included in their structure. In the plotted regime, the CI cannot be the best strategy.

The plot of SNR in the more noisy situation, $N_{B}=100$ and $K=10^{7}$, is shown in Fig.~\ref{FigSNRNS} (b). The regions of the double HD and the OPA receiver become wider, while that of the PC receiver goes narrower. This effect can be explained based on their measurement operator. As it was previously described in Sec.~\ref{SecSHD}, the measurement operators of the double HD and the OPA receiver include the return mode component $\hat{a}_{R}\hat{a}_{R}^{\dagger}$ of which expectation value depends on $N_{B}$, and thus, both numerator and denominator of SNR increase with growing $N_{B}$. However, since the measurement operator of the PC receiver does not contain the return mode component, only the denominator of SNR increases while the numerator is unchanged with increasing $N_{B}$. The CI cannot be the best in this regime as well.

Fig.~\ref{FigSNRD} shows the SNR of Gaussian QI with the double HD (red solid lines), the PC receiver (blue dashed lines), and the OPA receiver (orange dot-dashed lines) at $N_{S}=0.01$. The SNR of the coherent state CI is plotted as a benchmark as well (black dotted lines). Since the SNR of a QI system using single-mode is extremely small, $K$ should be large in order to amplify the SNR. In the both plots, we define $K=10^{7}$ to obtain $120$($\sim 20$dB) SNR at reflectivity $0.01$. 
Based on the experimental data\cite{Zhang2015,Barzanjeh2020}, we choose an intermediate value of $K=10^7$ which is expected to be realizable. Fig.~\ref{FigSNRD} (a) shows the SNR at $N_{B}=30$. The SNR with the double HD becomes the largest with increasing reflectivity. At $\kappa <0.0009$, the PC receiver shows the best performance in the QI, as shown in the inset. Fig.~\ref{FigSNRD} (b) shows the SNR at $N_{B}=100$, and it shows the same tendency with Fig.~\ref{FigSNRD} (a), except slightly lower SNR due to the large thermal noise. Since the double HD is less affected by the thermal noise than the PC receiver, the PC receiver shows the best performance at the smaller region $\kappa < 0.0003$.

We analyze the performance of a QI system using one of the three receivers at $>1\%$ reflectivity as well. Fig~\ref{FigSNRall} is a region plot of the best receiver in all the reflectivity regime and mean photon number of the signal $0\leq N_{S}\leq 1$ at $K=10^{7}$ and $N_{B}=30$. The PC receiver is the best choice at low $\kappa$ and large $N_{S}$. The double HD can be the best at low $\kappa$ and small $N_{S}$ and at high $\kappa$ and large $N_{S}$. Examples of the boundary reflectivity at $N_{S}=0.01$ and $0.001$ are drawn in Fig.~\ref{FigSNRall} as the black horizontal lines. At $N_{S}=0.01$, a QI system using the PC receiver is the best at $\kappa<0.0009$, the coherent state CI becomes the best at $\kappa >0.125$, and the double HD is the best in the middle range. In the case of $N_{S}=0.0001$, the double HD can outperform the others at $0.0001<\kappa<0.022$. If the reflectivity is larger than $0.022$, the CI becomes the best and the OPA receiver is the best otherwise as it was previously shown in Fig.~\ref{FigSNRNS} (a).

\section{Summary and discussion}\label{SecCon}

We proposed a new receiver setup for Gaussian QI. Performing double homodyne detection(HD) after combining the return and idler modes by a 50:50 beam splitter, we measured the mean square quadratures, $\braket{\hat{X}_{1}^{2}(0)+\hat{X}_{2}^{2}\left(\pi/2\right)}$. In comparison to the simple-structured receivers, such as the OPA and PC receivers, the double HD exhibited the enhanced target detection capability by the SNR. In low reflectivity regime ($\leq 0.01$) with low-power source (mean photon number of signal $\leq 0.02$), the double HD outperformed the other receivers mostly to detect the target. Also, due to its robustness against noise, the double HD will enhance the performance of a microwave QI system \cite{Barzanjeh2015,Luong2019,Barzanjeh2020}.

We analyzed these three receivers in various target reflectivity, source power, and noise level, and we found that the performance of a QI receiver depends on not only the structure of the QI receiver, but also the conditions of source and channel. Thus, a QI receiver should be chosen based on properties of the entire QI system such as power and bandwidth of the source. Our results can be a reference for selection of a QI receiver which gives the best performance in the QI system.

The SNRs with the OPA and PC receivers in Fig.~\ref{FigSNRD} (a) are not the same as the results in the previous study \cite{Guha2009} even under the same values of the parameters. The previous study assumed the condition $N_{B}\approx N_{B}/(1-\kappa)$, representing very noisy channel and very low reflectivity of the target, whereas there is no assumption in our analysis. Nonetheless, at extremely low reflectivity, the OPA and PC receivers in QI outperform the coherent CI, satisfying the assumption.

%%%%%%%%%%%%%%%%%%%%%%%%%%%%%%%%%%%%%%%%%%
\begin{acknowledgements}
This work was supported by a grant to the Quantum Standoff Sensing Defense-Specialized Project funded by the Defense Acquisition Program Administration and the Agency for Defense Development.
\end{acknowledgements}
%%%%%%%%%%%%%%%%%%%%%%%%%%%%%%%%%%%%%%%%%%

\begin{widetext}
\begin{appendix}
	\section{SNR equations}
	SNR can be calculated by using the measurement operators of Eqs.~(\ref{SHD}), (\ref{OPAM}), (\ref{PCM}) and the covariance matrices of Eqs.~(\ref{H0}), (\ref{H1}). In our noise channel, the SNR of QI with the dHD receiver is
	\begin{align}
		\text{SNR}_{\text{dHD}}^{(K)}=\frac{K\left[\kappa  (B-A)+2 C \sqrt{\kappa }\right]^2}{2 \left[\left|A (1+\kappa) +B(1-\kappa)-2 C \sqrt{\kappa }\right|+|A+B|\right]^2},
	\end{align}
	and that with the PC receiver is
	\begin{align}
		\text{SNR}_{\text{PC}}^{(K)}=\frac{K\kappa C^2  }{\left[\sqrt{\kappa  \left(A^2-A B+C^2\right)+A (B+2)+1}+\sqrt{A (B+2)+1}\right]^2},
	\end{align}
	and that with the OPA receiver is
	\begin{align}
		\text{SNR}_{\text{OPA}}^{(K)}=\frac{K\left[\kappa  (G-1) (A-B)+2 C \sqrt{\kappa G(G-1)}\right]^2}{2 \left(\sqrt{D_{0}}+\sqrt{D_{1}}\right)^2},
	\end{align}
	where $A=2N_{S}+1$, $B=2N_{B}+1$, $C=2\sqrt{N_{S}(N_{S}+1)}$, and
	\begin{align}
		\begin{split}
			D_{0}=&[A G+B (G-1)]^2-1,\\
			D_{1}=&A^2 [\kappa  (G-1)+G]^2-2 B (\kappa -1) (G-1) \left[A (\kappa  (G-1)+G)+2 C \sqrt{\kappa G (G-1)}\right]\\
			\quad &+4 A C \sqrt{\kappa G (G-1)} [\kappa  (G-1)+G]+B^2 (\kappa -1)^2 (G-1)^2+4 C^2 \kappa  (G-1) G-1.
		\end{split}
	\end{align}
The SNR of CI is too messy to write down here. So it can be derived by using the quantum Chernoff bound of single-mode Gaussian state\cite{Pirandola2008}.
\end{appendix}
\end{widetext}


\begin{thebibliography}{99}
	\bibitem{Bennett1984}
C. Bennett and G. Brassard, In {\it Proceedings of the IEEE International Conference on Computers, Systems and Signal Processing} (IEEE, 1984), p. 175--179.

\bibitem{Ekert1991}
A. Ekert, Phys. Rev. Lett. {\bf 67}, 661--663 (1991).

\bibitem{Feynman1982} % quantum computation
R. Feynman, Int. J. Theor. Phys. {\bf 21}, 467--488 (1982).

\bibitem{Lloyd2008} % Proposed the first QI
S. Lloyd, Science {\bf 321}, 1463--1465 (2008).

\bibitem{Shapiro2019} % The QI story
J. Shapiro, IEEE Aerosp. Electron. Syst. Mag. \textbf{35}(4), 8--20 (2020).

\bibitem{Tan2008} % Gaussian QI
S. Tan, B. Erkmen, V. Giovannetti, S. Guha, S. Lloyd, L. Maccone, S. Pirandola, and J. Shapiro, Phys. Rev. Lett. {\bf 101}, 253601 (2008).

\bibitem{SL09} J.H. Shapiro and S. Lloyd, New J. Phys. \textbf{11}, 063045 (2009).

\bibitem{Wilde2017}
M. M. Wilde, M. Tomamichel, S. Lloyd, and M. Berta, Phys. Rev. Lett. {\bf 119}, 120501 (2017).

\bibitem{Karsa2020} % QI with generic Gaussian source
A. Karsa, G. Spedalieri, Q. Zhuang, and S. Pirandola, Phys. Rev. Research {\bf 2}, 023414 (2020).

\bibitem{Lopaeva2013} % experimental QI
E. Lopaeva, I. Berchera, I. Degiovanni, S. Olivares, G. Brida, and M. Genovese, Phys. Rev. Lett. {\bf 110}, 153603 (2013).

\bibitem{Zhang2015} % experimental QI with OPA receiver
Z. Zhang, S. Mouradian, F. Wong, and J. Shapiro, Phys. Rev. Lett. {\bf 114}, 110506 (2015).

\bibitem{England2019} % experimental QI with correlated photon pairs
D. England, B. Balaji, and B. Sussman, Phys. Rev. A {\bf 99}, 023828 (2019).

\bibitem{Barzanjeh2015} % Microwave QI
S. Barzanjeh, S. Guha, C. Weedbrook, D. Vitali, J. Shapiro, and S. Pirandola, Phys. Rev. Lett. {\bf 114}, 080503 (2015)

\bibitem{Luong2019} % experimental demonstration of prototype microwave QI
D. Luong, C. Chang, A. Vadiraj, A. Damini, C. Wilson, and B. Balaji, IEEE Transactions on Aerospace and Electronic Systems (2019).

\bibitem{Barzanjeh2020} % Microwave quantum illumination experiment (OPA, PC receiver)
S. Barzanjeh, S. Pirandola, D. Vitali, and J. Fink, Sci. Adv. {\bf 6}, eabb0451 (2020).


%\bibitem{Tan} S.H. Tan, B.I. Erkmen, V. Giovannetti, S. Guha, S. Lloyd, L. Maccone, S. Pirandola, and J.H. Shapiro, \prl \textbf{101}, 253601 (2008). 

%\bibitem{Shapiro} J.H. Shapiro, arXiv:1910.12277.

%\bibitem{Genovese} E.D. Lopaeva, I. Ruo Berchera, I.P. Degiovanni, S. Olivares, G. Brida, and M. Genovese, \prl \textbf{110}, 153603 (2013).

%\bibitem{Guha}S. Guha and B.I. Erkmen, \pra \textbf{80}, 052310 (2009).

%\bibitem{Zheshen} Z. Zhang, S. Mouradian, F.N.C. Wong, and J.H. Shapiro, \prl \textbf{114}, 110506 (2015).

%\bibitem{Quntao} Q. Zhuang, Z. Zhang, and J.H. Shapiro, \prl \textbf{118}, 040801 (2017).




\bibitem{Devi}A.R. Usha Devi and A.K. Rajagopal, \pra  \textbf{ 79}, 062320 (2009).

\bibitem{Ragy} S. Ragy, I. Ruo Berchera, I. P. Degiovanni, S. Olivares, M. G. A. Paris, G. Adesso, and M. Genovese, J. Opt. Soc. Am. B \textbf{31}, 2045 (2014).

\bibitem{Zhang14} S.L. Zhang, J.S. Guo, W.S. Bao, J.H. Shi, C.H. Jin, X.B. Zou, and G.C. Guo, \pra  \textbf{ 89}, 062309 (2014).

%\bibitem{Shabir} S. Barzanjeh, S. Guha, C. Weedbrook, D. Vitali, J.H. Shapiro, and S. Pirandola, \prl \textbf{114}, 080503 (2015).

\bibitem{Sanz} M. Sanz, U. Las Heras, J.J. Garc\'ia-Ripoll, E. Solano, and R. Di Candia, \prl \textbf{ 118}, 070803 (2017).

\bibitem{Liu17} K. Liu, Q.-W. Zhang, Y.-J. Gu, and Q.-L. Li, \pra \textbf{ 95}, 042317 (2017).

\bibitem{Weedbrook}C. Weedbrook, S. Pirandola, J. Thompson, V. Vedral, and M. Gu, New. J. Phys.\textbf{ 18}, 043027 (2016).

\bibitem{Bradshaw}M. Bradshaw, S.M. Assad, J.Y. Haw, S.-H. Tan, P.K. Lam, and M. Gu, \pra \textbf{ 95}, 022333 (2017).

\bibitem{Zubairy}L. Fan and M.S. Zubairy, \pra \textbf{ 98}, 012319 (2018).

\bibitem{Stefano19} S. Pirandola, R. Laurenza, C. Lupo, and J.L. Pereira, npj Quantum Inf. \textbf{5}, 50 (2019).

\bibitem{Palma} G.De Palma and J. Borregaard, \pra \textbf{ 98}, 012101 (2018).

\bibitem{Yung} M.-H. Yung, F. Meng, X.-M. Zhang, and M.-J. Zhao, npj Quantum Inf. \textbf{6}, 75 (2020).

\bibitem{Ray} S. Ray, J. Schneeloch, C.C. Tison, and P.M. Alsing, \pra \textbf{ 100}, 012327 (2019).

\bibitem{Sun}W.-Z. Zhang, Y.-H. Ma, J.-F. Chen, and C.-P. Sun, New J. Phys. \textbf{22}, 013011 (2020).

\bibitem{Ranjith} R. Nair and M. Gu, Optica \textbf{7}, 771 (2020).

%\bibitem{Shabir19} S. Barzanjeh,  S. Pirandola, D. Vitali, and J.M. Fink, Sci. Adv. \textbf{6}, eabb0451 (2020).

\bibitem{Sandbo}C.W. Sandbo Chang, A. M. Vadiraj, J. Bourassa, B. Balaji, and C.M. Wilson, Appl. Phys. Lett. \textbf{114}, 112601 (2019).

%\bibitem{England}D.G. England, B. Balaji, and B.J. Sussman, \pra \textbf{99}, 023828 (2019).

\bibitem{Aguilar} G.H. Aguilar, M.A. de Souza, R.M. Gomes, J. Thompson, M. Gu, L. C. C\'eleri, and S. P. Walborn, \pra \textbf{ 99}, 053813 (2019).

\bibitem{Sussman} Y. Zhang, D. England, A. Nomerotski, P. Svihra, S. Ferrante, P. Hockett, and B. Sussman, Phys. Rev. A {\bf 101}, 053808 (2020).

\bibitem{Lee} S.-Y. Lee, Y.S. Ihn, and Z. Kim, arXiv:2004.09234 [quant-ph] (2020).

\bibitem{Guha2009} % OPA and PC receiver
S. Guha and B. Erkmen, Phys. Rev. A {\bf 80}, 052310 (2009).

\bibitem{Guha2009-1}
S. Guha, 2009 IEEE International Symposium on Information Theory, Seoul, pp. 963--967 (2009).

\bibitem{Zhuang2017} % FF-SFG
Q. Zhuang, Z. Zhang, and J. Shapiro, Phys. Rev. Lett. {\bf 118}, 040801 (2017).

\bibitem{Karsa2020-1}
A. Karsa and S. Pirandola, IEEE Aerospace and Electronic Systems Magazine {\bf 35}(11), pp. 22--29 (2020).

\bibitem{Audenaert2007} % Quantum Chernoff bound 
K. Audenaert, J. Calsamiglia, R. Mu{\~ n}oz-Tapia, E. Bagan, L. Masanes, A. Ac{\` i}n, and F. Verstraete, Phys. Rev. Lett. {\bf 98}, 160501 (2007).

\bibitem{Calsamiglia2008} % Quantum Chernoff bound for discrimination problem
J. Calsamiglia, R. Mu{\~ n}oz-Tapia, L. Masanes, A. Ac{\` i}n, and  E. Bagan, Phys. Rev. A {\bf 77}, 032311 (2008).

\bibitem{Pirandola2008} % Quantum Chernoff bound computation
S. Pirandola and S. Lloyd, Phys. Rev. A {\bf 78}, 012331 (2008).

\bibitem{Braunstein1990}
S. Braunstein, Phys. Rev. A {\bf 42}, 474 (1990).

%\bibitem{Butler1998} J. M. Butler, Ph.D. dissertation, University College London (1998).

	
\end{thebibliography}
\end{document}